\documentclass[
  aps,
  prl,
  reprint,
  amsmath,
  amssymb,
  footinbib,
  longbibliography,
  superscriptaddress
]{revtex4-2}

\usepackage{bm}
\usepackage{mathtools}
\usepackage[dvipsnames]{xcolor}
\usepackage{graphicx}

\usepackage[bookmarksnumbered=true,bookmarksopen=true]{hyperref}
 \hypersetup{colorlinks,%
             linkcolor=NavyBlue, %
             citecolor=PineGreen, %
             urlcolor=PineGreen}

\newcommand{\dd}{\mathrm{d}}
\newcommand{\ii}{\mathrm{i}}
\newcommand{\mnu}{\widetilde{\nu}}
\newcommand{\order}[1]{\mathcal{O}\!\left(#1\right)}

\newcommand{\ed}[1]{{\color{RoyalBlue}#1}}

\newcommand{\fr}[2]{\mbox{$\frac{\,{#1}\,}{#2}$}}

\begin{document}
\hbadness=10000

\title{Cosmological Collider Signals From a Triangle Loop}

\author{Zhehan Qin}
\email{qzh17@tsinghua.org.cn}
\affiliation{Department of Physics, Tsinghua University, Beijing 100084, China}
\affiliation{Department of Applied Mathematics and Theoretical Physics, University of Cambridge, Wilberforce Road, Cambridge, CB3 0WA, UK}

\author{Zhong-Zhi Xianyu}
\email{zxianyu@tsinghua.edu.cn}
\affiliation{Department of Physics, Tsinghua University, Beijing 100084, China}
\affiliation{Peng Huanwu Center for Fundamental Theory, Hefei, Anhui 230026, China}

\date{\today}

\begin{abstract}
Triangular loop signals appear frequently in cosmological collider models but their analytical forms remain unknown due to multiple technical challenges. Here, we determine for the first time the leading analytical cosmological collider signals generated by a massive scalar triangle loop in inflationary bispectrum and trispectrum. Combining a directional cutting rule with partial Mellin--Barnes representations, we obtain both the local and nonlocal trispectrum signals in a soft limit and derive the leading signal in the squeezed bispectrum. Remarkably, all these signals are reproduced by the corresponding bubble diagram with a single effective pinched coupling. Signals beyond leading order and cases with arbitrary internal masses, spins, and interactions can be generated within the same framework. Our approach also provides a systematic route to extracting analytical loop signals from general cosmological correlators. 
\end{abstract}

\maketitle

\emph{Introduction.---} 
The current and future observations of large-scale nonuniformity of our universe bring exciting opportunities to probe fundamental particles and their interactions at the inflation scale through the cosmological collider (CC) observables \cite{Chen:2009zp,Baumann:2011nk,Noumi:2012vr,Arkani-Hamed:2015bza}. Studies in the past decade have identified many promising particle and cosmological models that can generate visible signals in generic parameter space. Thus, a major task for current study is to connect these model predictions with observational data, which has received considerable attention recently \cite{Cabass:2024wob,Goldstein:2024bky,Sohn:2024xzd,Suman:2025vuf,Suman:2025tpv,Philcox:2025wts,Bao:2025onc,Anbajagane:2025uro,Green:2026yev,Kumar:2026ogn,Kumar:2026dih,Cassem:2026ygh,Philcox:2026rpn,Philcox:2026tjj}.

To bridge the gap between models and data, fast and high-quality templates from realistic models are essential. Nevertheless, generating templates is generally difficult and slow due to many technical challenges and has become a main bottleneck for current studies. Many activities in recent years have made major progress in this direction, especially for tree-level processes where various methods were developed and many results obtained \cite{Arkani-Hamed:2018kmz,Baumann:2019oyu,Pimentel:2022fsc,Jazayeri:2022kjy,Qin:2022fbv,Qin:2023ejc,Liu:2024str,Qin:2025xct,Sleight:2019mgd,Sleight:2019hfp,Sleight:2020obc,Qin:2022lva,Qin:2022fbv,Xianyu:2023ytd,Qin:2024gtr,Liu:2024xyi,Chen:2024glu,Werth:2024mjg,Xianyu:2025lbk,Cespedes:2025ple,Wang:2025qfh,Baumann:2026atn,Arundine:2026myr,Belrhali:2026ktb,Belrhali:2026rkn,Belrhali:2026jqe,Belrhali:2026ygh,Belrhali:2026act,Belrhali:2026uxn,Grafe:2026avi,Wang:2026lff,Pinol:2026xnl,Huenupi:2026aqc,Huenupi:2026abj}. 

On the other hand, loop signals are ubiquitous in CC physics. Many model studies suggest that loop signals are often leading when massive states carry conserved charges of unbroken symmetries or half-integer spins. Such signals often arise from all possible 1-loop topologies, including the bubble and triangle for bispectrum, and also the box for the trispectrum \cite{Chen:2016nrs,Chen:2016uwp,Chen:2016hrz,Lu:2019tjj,Hook:2019zxa,Hook:2019vcn,Kumar:2018jxz,Lu:2021gso,Cui:2021iie,Bodas:2025vpb,Aoki:2026olh,You:2026xoq,Chen:2018xck,Wang:2019gbi,Wang:2020ioa,Tong:2022cdz}. In many models, these loop signals also receive natural enhancement that is absent at tree level \cite{Chen:2018xck,Wang:2019gbi,Wang:2020ioa,Sou:2021juh,Tong:2022cdz,An:2025mdb}.  However, templates for loop signals remain very difficult; existing full loop results are largely confined to the two-vertex bubble/banana families \cite{Xianyu:2022jwk,Liu:2024xyi,Qin:2024gtr,Zhang:2025nzd,Grafe:2026qsm,Aoki:2026vbc,Liu:2026jzn} or conformally coupled scalars \cite{Pimentel:2026kqc,Henn:2026lfz}. Therefore, it is of vital importance to gain more analytic control over massive loops beyond the simple bubble.
 
\begin{figure}[t]
  \centering
  \includegraphics[width=1.0\columnwidth]{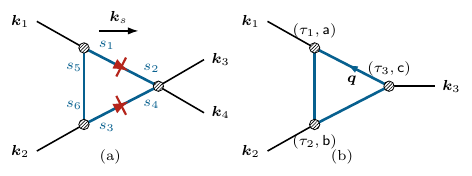}\\[-0.5mm]
   \caption{Triangle topologies and directional cut.
Blue lines denote massive $\sigma$ propagators and black lines external inflatons.
Hatched vertices indicate Schwinger--Keldysh (SK) indices summed over $\pm$.
In (a), the red transverse marks and arrowheads indicate the directional cut, with the arrows pointing toward the low-energy vertex.
The Mellin endpoint variables are shown in (a), while (b) displays the vertex times and their SK indices; the same assignments apply to both diagrams.
The arrow above (a) denotes the soft channel momentum $\bm k_s$, while the arrow on the upper propagator in (b) fixes the loop momentum $\bm q$.}
  \label{fig:triangles}
\end{figure}

In this \emph{Letter}, we report new analytical progress in understanding triangular loop signals.  Using partial Mellin--Barnes (PMB) representations~\cite{Qin:2022lva,Qin:2022fbv},
we obtain the complete leading  signals of the
four- and three-point one-loop triangles shown in Fig.~\ref{fig:triangles},
in the relevant soft limits.
We further validate our analytical results against direct numerical evaluations of the original loop integrals, finding excellent agreement. A key ingredient is a directional cutting rule, which isolates the
signal-carrying part of the full correlator and substantially simplifies
its computation.
Remarkably, the newly obtained local signal dominates the four-point
triangle contribution in the further collapsed regime.
Moreover, the complete leading triangle signal admits a simple reduction to the corresponding bubble signal through an effective pinched coupling.
The framework also allows systematic extensions to subleading orders
and more general internal masses, spins, and interactions. 

Despite the many technicalities involved, simple physical intuitions go into our signal extraction: CC
signals arise from on-shell massive propagation and can therefore be
isolated by cutting the relevant internal lines~\cite{Tong:2021wai,Qin:2023bjk,Qin:2023nhv,Ema:2024hkj}.
Previous applications of this idea obtained the leading four-point nonlocal
signals \footnote{For a four-point function, the signal splits into two classes: the nonlocal signal, nonanalytic in the exchange momentum, and the local signal, analytic in it but nonanalytic in another momentum ratio.} of triangle and box diagrams, and showed that these signals can equivalently be reproduced by the corresponding bubble diagrams with effective pinched couplings~\cite{Qin:2023bjk}.
Here, we go beyond the nonlocal signal and show that the directional cut captures both the nonlocal and local contributions to four-point signals and applies equally to the bispectrum triangle signal.
More broadly, our results establish this cutting procedure as a systematic tool for extracting complete  signals at loop level.

\emph{Setup and conventions.---}
We use spatially flat de Sitter metric
$\dd s^2=a^2(\tau)(-\dd\tau^2+\dd\bm x^2)$, where
$a(\tau)=-1/(H\tau)$, and set $H\equiv1$ throughout. Bold symbols $\bm k$
denote three-dimensional momenta, with $k\equiv|\bm k|$ their magnitudes.
We abbreviate indexed sums as
$k_{ij\cdots}\equiv k_i+k_j+\cdots$ and
$s_{ij\cdots}\equiv s_i+s_j+\cdots$, except for the step function $\theta_{ij}\equiv\theta(\tau_i-\tau_j)$.
We use $\int_\tau\equiv\int_{-\infty}^0\dd\tau$ and
$\int_s\equiv\int_{-\ii\infty}^{+\ii\infty}\dd s/(2\pi\ii)$.
The real parts of the Mellin contours are left implicit: each contour runs
vertically within a fundamental strip separating the left- and right-pole
sequences.

We consider an inflaton fluctuation $\phi$ interacting with a massive scalar
$\sigma$ through the interaction
\begin{equation}
 \mathcal L_{\mathrm{int}}
 =\frac12 a^3\phi'\sigma^2+\frac14a^2\phi'^2\sigma^2,
 \label{eq:interaction}
\end{equation}
with the dimensionful couplings omitted. More general couplings can be handled similarly. Two cubic vertices and one
quartic vertex generate the four-point triangle in
Fig.~\ref{fig:triangles}(a), while three cubic vertices generate the three-point
triangle in Fig.~\ref{fig:triangles}(b). With a prime denoting the removal of
the overall factor $(2\pi)^3\delta^3(\sum_i\bm k_i)$ from an equal-time
correlator, we write the $s$-channel four-point correlator (trispectrum) and
the three-point correlator (bispectrum) as
$\mathfrak T\equiv
\langle\phi_{\bm k_1}\phi_{\bm k_2}\phi_{\bm k_3}\phi_{\bm k_4}\rangle'_s$
and $\mathfrak B\equiv
\langle\phi_{\bm k_1}\phi_{\bm k_2}\phi_{\bm k_3}\rangle'$, respectively,
where $\bm k_s\equiv\bm k_1+\bm k_2$ in the former case.

Assuming the Bunch--Davies vacuum,
the mode functions for $\phi$ and $\sigma$ are, respectively,
\begin{align}
 \phi_k(\tau)&=\frac{1}{\sqrt{2k^3}}
 (1+\ii k\tau)e^{-\ii k\tau},
 \label{eq:massless-mode}\\
 \sigma_k(\tau)&=\frac{\sqrt{\pi}}{2}
 e^{-\pi\mnu/2}(-\tau)^{3/2}
 \mathrm H_{\ii\mnu}^{(1)}(-k\tau),
 \label{eq:massive-mode}
\end{align}
where $\mnu\equiv\sqrt{m^2-9/4}$. For definiteness, we take $\sigma$ to lie in the principal series, so
that $\mnu$ is real; the complementary-series expressions follow by analytic
continuation, $\mnu\to\pm\ii\sqrt{9/4-m^2}$.
We follow the conventions of Ref.~\cite{Chen:2017ryl} for the bulk-to-boundary propagator
$K_\pm$ and the bulk-to-bulk propagator $D_{\mathsf a\mathsf b}$;
the explicit expressions are collected in the Supplemental
Material~\cite{SupplementalMaterial}.

Let $\{\tau\}\equiv(\tau_1,\tau_2,\tau_3)$. Following the Schwinger--Keldysh (SK) rules in Ref.~\cite{Chen:2017ryl}, the trispectrum $\mathfrak T$ shown in Fig.~\ref{fig:triangles}(a) is given by~\footnote{In this work we focus on the ordered $s$-channel contribution shown in Fig.~\ref{fig:triangles}(a), without summing over permutations of the
external legs. For identical external fields, the corresponding
permutations should in general be included, and some crossed permutations
could also generate oscillatory signals. In the configuration considered here, they approach squeezed-triangle kinematics and can be treated analogously to the three-point case. A systematic
treatment of the full permutation sum is discussed in a companion work.}
\begin{align}
 \mathfrak T={}&
 \sum_{\mathsf a,\mathsf b,\mathsf c=\pm}
 (-\ii\mathsf a\mathsf b\mathsf c)\int_{\{\tau\}}
 a^3(\tau_1)a^3(\tau_2)a^2(\tau_3)\mathbb J_{\mathsf a\mathsf b\mathsf c}(\bm k_s;\{\tau\})
 \nonumber\\
 &\times K'_{\mathsf a}(k_1,\tau_1)K'_{\mathsf b}(k_2,\tau_2) K'_{\mathsf c}(k_3,\tau_3)K'_{\mathsf c}(k_4,\tau_3),
 \label{eq_Ts}
\end{align}
 while the bispectrum $\mathfrak B$ in Fig.~\ref{fig:triangles}(b) is
 \begin{align}
 \mathfrak B={}&
 \sum_{\mathsf a,\mathsf b,\mathsf c=\pm}
 (-\ii\mathsf a\mathsf b\mathsf c)\int_{\{\tau\}}
 \prod_{j=1}^3a^3(\tau_j)\mathbb J_{\mathsf a\mathsf b\mathsf c}(-\bm k_3;\{\tau\})\nonumber\\
 &\times K'_{\mathsf a}(k_1,\tau_1)K'_{\mathsf b}(k_2,\tau_2)K'_{\mathsf c}(k_3,\tau_3),
 \label{eq_B}
\end{align}
where the common triangle loop kernel is defined by
\begin{align}
 \mathbb J_{\mathsf a\mathsf b\mathsf c}(\bm k;\{\tau\})
 \equiv{}&\int\frac{\dd^3\bm q}{(2\pi)^3}
 D_{\mathsf a\mathsf c}(q;\tau_1,\tau_3)
 D_{\mathsf b\mathsf c}(|\bm q+\bm k|;\tau_2,\tau_3)
 \nonumber\\
 &\times
 D_{\mathsf a\mathsf b}(|\bm q+\bm k_1|;\tau_1,\tau_2).
 \label{eq_J}
\end{align}
Here the dependence on the fixed momentum $\bm k_1$ is left implicit.

The two correlators in Eq.~\eqref{eq_Ts} and Eq.~\eqref{eq_B} are governed by the same triangle kernel in Eq.~\eqref{eq_J}, and have the same
interaction vertices at $\tau_{1,2}$. Their time integrands differ only in the power
of $\tau_3$ arising from the external insertion. This
motivates the following master triangle integral
\begin{align}
 \mathcal T^p(\bm k,E)\equiv{}&
 \sum_{\mathsf a,\mathsf b,\mathsf c=\pm}
 (-\ii\mathsf a\mathsf b\mathsf c)
 \int_{\{\tau\}}\frac{(-\tau_3)^p}{(-\tau_1)^2(-\tau_2)^2}
 \nonumber\\
 &\times
 e^{\ii\mathsf a k_1\tau_1+\ii\mathsf b k_2\tau_2
   +\ii\mathsf c E\tau_3}
 \mathbb J_{\mathsf a\mathsf b\mathsf c}(\bm k;\{\tau\}).
 \label{eq:master-triangle-integral}
\end{align}
Here $\bm k$ is the momentum flowing out from the vertex at $\tau_3$,
and $E$ is the total external energy attached to that vertex.  Direct
comparison with Eq.~\eqref{eq_Ts} and Eq.~\eqref{eq_B} gives
\begin{align}
 \mathfrak T
 =\frac{\mathcal T^0(\bm k_s,k_{34})}{16k_1k_2k_3k_4},\quad
 \mathfrak B=-\frac{\mathcal T^{-2}(-\bm k_3,k_3)}{8k_1k_2k_3},
 \label{eq:triangle-master-cases}
\end{align}
where the choices $p=0$ and $p=-2$ encode the powers of $\tau_3$ generated by
the quartic and cubic insertions at $\tau_3$, respectively. In the remainder
of the \emph{Letter}, we compute the signal parts of these reduced master integrals;
the external-line factors are restored using Eq.~\eqref{eq:triangle-master-cases}.
 
\emph{PMB representation and directional cutting.---}
PMB representations provide a useful framework for computing cosmological correlators: they factorize the time integrals from the loop-momentum integrals and make the relevant nonanalytic structure manifest \cite{Qin:2023nhv,Qin:2023bjk,Qin:2024gtr}. We use Mellin--Barnes representations of massive mode functions \cite{Sleight:2019mgd,Sleight:2019hfp}, while leaving the step functions untouched, such that
\begin{align}
 \sigma_k(\tau)
 =& -\frac{\ii}{2\sqrt{\pi}}(-\tau)^{3/2}
 \int_s
 e^{\ii\pi s}\left(\frac{-k\tau}{2}\right)^{-2s} \nonumber\\
 &\times
 \Gamma\!\left(s+\frac{\ii\mnu}{2}\right)
 \Gamma\!\left(s-\frac{\ii\mnu}{2}\right).
 \end{align}
The phases and step functions can then be collected into the nesting factors
\begin{align}
 &\mathcal N_{\mathsf a,-\mathsf a}(s_1,s_2;\tau_1,\tau_2)\equiv \mathcal N_{\mathsf a,-\mathsf a}(s_1,s_2) = \frac{2^{2s_{12}}}{4\pi}
 e^{-\ii\pi\mathsf a(s_1-s_2)},\nonumber\\
 &\mathcal N_{\mathsf a\mathsf a}(s_1,s_2;\tau_1,\tau_2)=\theta_{12}\,
 \mathcal N_{-\mathsf a,\mathsf a}(s_1,s_2)
 +\theta_{21}\,
 \mathcal N_{\mathsf a,-\mathsf a}(s_1,s_2),
 \label{eq:Mellin-kernel}
\end{align}
so that the bulk-to-bulk propagator takes the form
\begin{align}
 \MoveEqLeft[1]
 D_{\mathsf a\mathsf b}(k;\tau_1,\tau_2)
 =
 \int_{s_{1,2}}
 \mathcal N_{\mathsf a\mathsf b}(s_1,s_2;\tau_1,\tau_2)k^{-2s_{12}}
 \nonumber\\[-1mm]
 \MoveEqLeft[1]
 \times 
 \prod_{j=1}^2\left[
 (-\tau_j)^{3/2-2s_j}
 \Gamma\!\left(s_j+\frac{\ii\mnu}{2}\right)
 \Gamma\!\left(s_j-\frac{\ii\mnu}{2}\right)
 \right].
 \label{eq_DinPMB}
\end{align}

To isolate a kinematic nonanalyticity, the appropriate cut is determined by the corresponding on-shell process together with the hierarchy of external energies: the former specifies which internal lines are cut, while the latter fixes the direction of the cut. For the four-point function, we always take the hierarchical configuration $k_{12}\gg k_{34}\geq k_s$, with $k_s/k_{34}\leq1$ held fixed. After factoring out the overall homogeneous scaling, this limit can equivalently be viewed as the hard limit $k_{12}\to\infty$ at fixed $k_{34}$ and $k_s$. For the three-point function, we take the squeezed limit $k_{12}\gg k_3$. In both cases, the vertex at $\tau_3$ carries the lower external energy and therefore fixes the cut direction. The directional cutting rule, derived in the Supplemental Material~\cite{SupplementalMaterial}, then replaces the two lines ending at this vertex by directional cut propagators,
\begin{align}
 D_{\mathsf{a}\mathsf{c}}(q;\tau_1,\tau_3)
 &D_{\mathsf{b}\mathsf{c}}(|\bm q+\bm k|;\tau_2,\tau_3)
 \ \longrightarrow \nonumber\\
 &D_{-\mathsf{c},\mathsf{c}}(q;\tau_1,\tau_3)
 D_{-\mathsf{c},\mathsf{c}}(|\bm q+\bm k|;\tau_2,\tau_3),
 \label{eq:cutting}
\end{align}
and thereby isolates the contributions capable of generating CC signals.

In the PMB representation, the cut further simplifies the time integrals by removing time ordering across the cut.
More explicitly, we define the cut master integral $\mathcal T_{\rm cut}^p$
from Eq.~\eqref{eq:master-triangle-integral} by replacing the two propagators attached to the $\tau_3$ vertex according to Eq.~\eqref{eq:cutting}.
Let $\{s\}\equiv(s_1,\cdots, s_6)$. Inserting the PMB representations of the massive mode functions then gives
the factorized form
\begin{align}
 \mathcal T_{\rm cut}^p(\bm k,E)
 ={}&\int_{\{s\}}
 \mathbb T_\mathrm{cut}^p(\{s\};k_1,k_2,E)
 \mathbb L(\{s\};\bm k,\bm k_1)\nonumber\\[-1mm]
 &\times\prod_{j=1}^6
 \Gamma\!\left(s_j+\frac{\ii\mnu}{2}\right)
 \Gamma\!\left(s_j-\frac{\ii\mnu}{2}\right)\ed{.}
 \label{eq:cut-factorization}
\end{align}
In particular, the four- and three-point cases are obtained from the cut master integral by setting $(p,\bm k,E) = (0,\bm k_s, k_{34})$ and $(p,\bm k,E) = (-2,-\bm k_3, k_3)$, respectively.
Performing the now-factorized $\tau_3$
 integral and the $\mathsf c$ sum using the nesting factor in Eq.~\eqref{eq:Mellin-kernel}, we obtain the cut time part,
\begin{align}
 \MoveEqLeft[1]
 \mathbb T_\mathrm{cut}^p
 \equiv-\frac{2^{2s_{1234}}\sin\!\left[\pi\left(s_{13}-\frac{p}{2}\right)\right]}{8\pi^2}
 \Gamma(p+4-2s_{24})\nonumber\\[-1mm]
 \MoveEqLeft[1]
 \times E^{-p-4+2s_{24}}
 \sum_{\mathsf a,\mathsf b=\pm}(-\mathsf a\mathsf b)
 \int_{\tau_{1,2}}\mathcal N_{\mathsf a\mathsf b}(s_5,s_6;\tau_1,\tau_2)\nonumber\\[-1mm]
 \MoveEqLeft[1]
 \times
 (-\tau_1)^{1-2s_{15}}(-\tau_2)^{1-2s_{36}}e^{\ii\mathsf a k_1\tau_1+\ii\mathsf b k_2\tau_2}.
 \label{eq:time-part}
\end{align}
Although the remaining $\tau_{1,2}$
 integrals can be performed explicitly, we keep them in the present form, as they will later recombine directly into a tree-level single-exchange seed.

 The loop part is a massless triangle integral, which admits a triple-$K$ representation~\cite{Bzowski:2013sza}
\begin{align}
 \mathbb L\equiv\int\frac{\dd^3\bm q}{(2\pi)^3}
 q^{-2s_{12}}|\bm q+\bm k|^{-2s_{34}}
 |\bm q+\bm k_1|^{-2s_{56}}.
 \label{eq:loop-part}
\end{align}
In both applications, the relevant hierarchy is $k_{12}\gg E\geq k$, and hence $k\ll k_{12}$. Since $\bm k=\bm k_1+\bm k_2$, momentum conservation then gives $|k_1-k_2|\leq k$, and hence $k_{1,2}=k_{12}/2+\order{k}$. Accordingly, we henceforth set $k_1=k_2=k_{12}/2$, with corrections in $k/k_{12}$ understood. For this isosceles configuration, the cut time integral scales as $\mathbb T^p_{\rm cut} \propto E^{-p-4+2s_{24}}k_1^{-4+2s_{1356}}$. The loop integral was evaluated in Ref.~\cite{Qin:2023bjk}; its leading small-$k/k_{12}$ form is
\begin{align}
 \mathbb L\simeq{}&
 \frac{k_1^{-2s_{56}}k^{3-2s_{1234}}}{(4\pi)^{\frac32}}
 \frac{\Gamma(s_{1234}-\frac32)
 \Gamma(\frac32-s_{12})\Gamma(\frac32-s_{34})}
 {\Gamma(3-s_{1234})\Gamma(s_{12})\Gamma(s_{34})}
 \nonumber\\
 &+\frac{k_1^{3-2s_{123456}}}{(4\pi)^{\frac32}}
 \frac{\Gamma(s_{123456}-\frac32)
 \Gamma(\frac32-s_{1234})\Gamma(\frac32-s_{56})}
 {\Gamma(3-s_{123456})\Gamma(s_{1234})\Gamma(s_{56})}.
 \label{eq:isosceles-loop}
\end{align}
The full expression for $\mathbb L$ is given in the Supplemental Material~\cite{SupplementalMaterial}.

The two terms in Eq.~\eqref{eq:isosceles-loop}, denoted by $\mathbb L_1$ and
$\mathbb L_2$, arise from the soft and hard loop-momentum regions,
respectively~\cite{Beneke:1997zp}. For $\mathbb L_1$, $q\sim k\ll k_{12}$ and
expanding the hard propagator reduces the triangle to a bubble integral. For
$\mathbb L_2$, $q\sim k_{12}$ and one instead expands in the soft external
momentum $k$. Thus, $\mathbb L_1$ can be nonanalytic in $k$, whereas
$\mathbb L_2$ must be analytic in $k$.

With both time and loop integrals understood in the Mellin space, it remains to finish the Mellin integrals over $s_{1,\cdots,6}$, which can be executed by carefully choosing contours and picking up appropriate poles, which we detail in the End Matter. In short, the leading signal comes solely from the same-sign spectral poles (\ref{eq_LeadingSignalPoles}) which fix $s_{1,3}$; The factorization of the signal expresses the time integral and $s_{5,6}$-integral in terms of a single-exchange tree seed $\mathcal I_{\mnu}^{p_1,
 p_2}(1,1)$ given in (\ref{eq:seed-special}); Also denoting the result of the remaining $s_{2,4}$ integrations by
$\mathcal M_{\mathsf c}^p(k/E)$, whose explicit hypergeometric form is
given in the Supplemental Material~\cite{SupplementalMaterial}, the
leading master signal factorizes as
\begin{align}
 &\mathcal T_{\rm sig}^p(\bm k,E)
 =
 \frac{k^3/(k_1^4E^{p+4})}{64\pi^{7/2}}
 \sum_{\mathsf c=\pm1}
 \left(\frac{k}{2k_1}\right)^{2\ii\mathsf c\mnu}
 \mathcal M_{\mathsf c}^p\!\left(\frac{k}{E}\right)\nonumber\\
 &\times \sin\!\left[
 \pi\!\left(\ii\mathsf c\mnu+\fr p2\right)\right]
 \Gamma^2(-\ii\mathsf c\mnu)
 \,\mathcal I_{\mnu}^{-\frac12+\ii\mathsf c\mnu,
 -\frac12+\ii\mathsf c\mnu}(1,1).
 \label{eq:master-signal}
\end{align}

\begin{figure}[t]
  \centering
    \includegraphics[width=1.0\columnwidth]{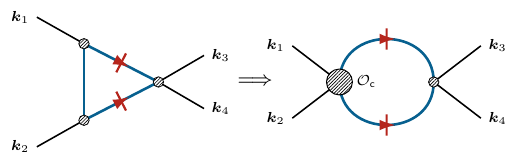}\\[-0.5mm]
\caption{
Pinched representation of the leading triangle signal.
For the four-point case shown here, the hard propagator and its two adjacent cubic vertices in the directionally cut triangle are pinched into the effective quartic operator
$\mathcal O_{\mathsf c}=a^2\mathcal C_{\mathsf c}\phi'^2\sigma^2/4$,
yielding the corresponding cut bubble diagram, while the two cut massive propagators remain unchanged.
In the hierarchy $k_1\simeq k_2\gg k_{34}\ge k_s$, this reproduces the complete leading four-point signal, including both the local and nonlocal branches.
The same pinched procedure also applies to the complete leading three-point signal.
}
  \label{fig:pinch-procedure}
\end{figure}

In particular, the tree-level seed $\mathcal{I}$ in
Eq.~\eqref{eq:master-signal} represents the hard subgraph formed by the
uncut propagator and its two adjacent vertices, including the additional
power-law factors $(-\tau_{1,2})^{3/2+\ii\mathsf c\mnu}$ inherited from the cut lines.
Pinching the hard propagator, as illustrated in Fig.~\ref{fig:pinch-procedure}, replaces this subgraph by an effective
quartic vertex $\mathcal O_\mathsf c = a^2\mathcal C_{\mathsf c}\phi'^2\sigma^2/4$, whose
coefficient defines the pinched coupling
\begin{equation}
  \mathcal C_{\mathsf c}(\mnu)
  =-\frac{4(2+\ii\mathsf c\mnu)}
  {(3+2\ii\mathsf c\mnu)^2(1+\ii\mathsf c\mnu)},
  \qquad \mathsf c=\pm1 .
\end{equation}
This is precisely the coupling obtained in
Ref.~\cite{Qin:2023bjk} for the leading nonlocal four-point triangle signal.
Remarkably, Eq.~\eqref{eq:master-signal} shows that the same pinched
coupling applies to both the local and nonlocal signals.
The leading triangle signal can therefore be reproduced by the
corresponding bubble diagram with the same effective coupling, for both the
three- and four-point functions. Also, the pinched coupling $\mathcal{C}_{\mathsf c}\sim 1/\mnu^2$ for $\mnu\gg 1$ which fits nicely with the EFT intuition in the large mass limit.

\begin{figure}[t]
  \centering
  \includegraphics[width=0.54\columnwidth]{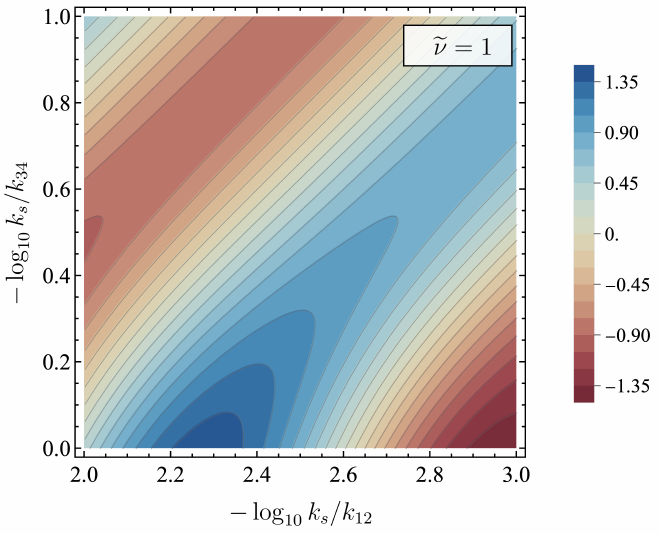}
  \hfill\includegraphics[width=0.44\columnwidth]{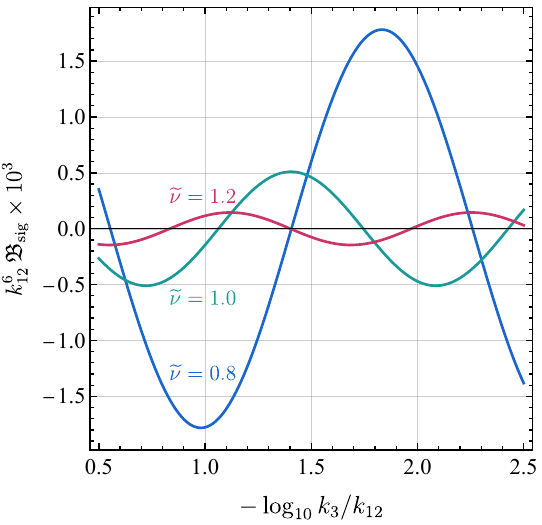}\\[-0.5mm]
  \caption{Rescaled leading signals from the massive triangle loop.
Left: Color map of
$10^3 k_{12}^6 k_3 k_4 k_{34}\mathfrak T_{\rm sig}$,
with $\mathfrak T_{\rm sig}=\mathfrak T_{\rm NS}+\mathfrak T_{\rm LS}$
as in Eq.~\eqref{eq:trispectrum-split},
as a function of $-\log_{10}(k_s/k_{12})$ and
$-\log_{10}(k_s/k_{34})$, using the leading result in
$k_{34}/k_{12}\ll1$ with $k_s/k_{34}$ fixed, for $\mnu=1$.
The finite-$k_s/k_{34}$ dependence deforms the oscillatory pattern
away from the further collapsed limit, while the total signal remains
finite at the folded boundary $k_s=k_{34}$.
Right: $10^3 k_{12}^6\mathfrak B_{\rm sig}$ from
Eq.~\eqref{eq:bispectrum-signal}, using the leading squeezed result in
$k_3/k_{12}\ll1$, plotted against $-\log_{10}(k_3/k_{12})$
for $\mnu=0.8,1.0,1.2$.}
  \label{fig:signals}
\end{figure}

\emph{Four-point signal.---}
We first specialize the master result to the four-point function by
setting $p=0$, $k=k_s$, and $E=k_{34}$.  Restoring the external-line
factors using Eq.~\eqref{eq:triangle-master-cases}, the leading
trispectrum signal separates as
\begin{equation}
 \mathfrak T_{\rm sig}
 =
 \mathfrak T_{\rm NS}
 +
 \mathfrak T_{\rm LS}.
 \label{eq:trispectrum-split}
\end{equation}
The nonlocal signal is
\begin{align}
 {}&\mathfrak T_{\rm NS}
 =
 \frac{k_s^3\sinh(\pi\mnu)}
 {64\pi^{5/2}k_{12}^6k_3k_4k_{34}^4}
 \sum_{\mathsf c=\pm1}
 \ii\mathsf c
 \left(
 \frac{k_s^2}{4k_{12}k_{34}}
 \right)^{2\ii\mathsf c\mnu}
 \nonumber\\[-1mm]
 &\times
 \frac{
 (2+\ii\mathsf c\mnu)
 \Gamma(3+2\ii\mathsf c\mnu)
 }
 {\Gamma(2+\ii\mathsf c\mnu)}\Gamma\!\left(-\frac32-2\ii\mathsf c\mnu\right)\Gamma(-\ii\mathsf c\mnu)
 \nonumber\\[-1mm]
 &\times
 \Gamma^2\!\left(\frac32+\ii\mathsf c\mnu\right)
 {}_2\mathrm F_1\!\left[
 \begin{matrix}
 2+\ii\mathsf c\mnu,\;
 \frac52+\ii\mathsf c\mnu\\
 \frac52+2\ii\mathsf c\mnu
 \end{matrix}
 \middle|\frac{k_s^2}{k_{34}^2}
 \right],
 \label{eq:trispectrum-nonlocal}
\end{align}
while the local signal is
\begin{align}
 {}&\mathfrak T_{\rm LS}
 =
 \frac{\sinh(\pi\mnu)}
 {8\pi^{5/2}k_{12}^6k_3k_4k_{34}}
 \sum_{\mathsf c=\pm1}
 \ii\mathsf c
 \left(
 \frac{k_{34}}{k_{12}}
 \right)^{2\ii\mathsf c\mnu}
 \nonumber\\[-1mm]
 &\times
 \frac{
 (2+\ii\mathsf c\mnu)
 \Gamma(1-2\ii\mathsf c\mnu)}
 {(3+2\ii\mathsf c\mnu)
 \Gamma(2+\ii\mathsf c\mnu)}\Gamma\!\left(\frac32+2\ii\mathsf c\mnu\right) \Gamma(-\ii\mathsf c\mnu)
 \nonumber\\[-1mm]
 &\times
 \Gamma^2\!\left(\frac32+\ii\mathsf c\mnu\right)
 {}_2\mathrm F_1\!\left[
 \begin{matrix}
 \frac12-\ii\mathsf c\mnu,\;
 1-\ii\mathsf c\mnu\\
 -\frac12-2\ii\mathsf c\mnu
 \end{matrix}
 \middle|\frac{k_s^2}{k_{34}^2}
 \right].
 \label{eq:trispectrum-local}
\end{align}
At this leading order in the soft expansion $k_{34}/k_{12}\ll 1$, the
hypergeometric functions retain the full dependence on
$k_s/k_{34}$ and encode the deformation of the signal away
from the further collapsed limit $k_s/k_{34}\to0$, as shown
in the left panel of Fig.~\ref{fig:signals}.

The relative importance of the two signals becomes particularly
transparent in the further collapsed limit $k_s\ll k_{34}$, where both
hypergeometric functions approach unity.  Comparing the envelopes of
the two oscillatory branches gives
\begin{align}
 \frac{|\mathfrak T_{{\rm LS},\mathsf c}|}
 {|\mathfrak T_{{\rm NS},\mathsf c}|}
 ={}&
 8\left(\frac{k_{34}}{k_s}\right)^3
 \left|
 \frac{
 \Gamma(1-2\ii\mathsf c\mnu)
 \Gamma\!\left(\frac32+2\ii\mathsf c\mnu\right)}
 {
 \Gamma(4+2\ii\mathsf c\mnu)
 \Gamma\!\left(-\frac32-2\ii\mathsf c\mnu\right)}
 \right|.
 \label{eq:local-nonlocal-ratio}
\end{align}
The mass-dependent prefactor in
Eq.~\eqref{eq:local-nonlocal-ratio} is finite and of order unity
for $\mnu\sim\mathcal{O}(1)$. Thus, in the collapsed regime, the local signal is parametrically larger than the nonlocal signal by a factor of order $(k_{34}/k_s)^3$. This is also visible in the left panel of
Fig.~\ref{fig:signals}: the stronger variation along the
anti-diagonal direction reflects the local oscillatory factor
$(k_{34}/k_{12})^{\pm 2\ii\mnu}$, whereas the weaker diagonal
variation is associated with the nonlocal oscillatory factor
$[k_s^2/(4k_{12}k_{34})]^{\pm 2\ii\mnu}$. Furthermore, the remaining analytic background $\mathfrak T_{\rm BG}$ generated by the same interaction
is expected to scale as a $\phi'^4$ contact contribution,
$\mathfrak T_{\rm BG}\sim 1/(k_1k_2k_3k_4k_T^5)
\simeq 1/(k_{12}^7k_3k_4)$.
Comparing it with Eq.~\eqref{eq:trispectrum-local} gives
$|\mathfrak T_{\rm LS}/\mathfrak T_{\rm BG}|\sim k_{12}/k_{34}\gg1$, and thus the local signal
parametrically dominates the full triangle correlator in the hierarchical
collapsed regime $k_s\ll k_{34}\ll k_{12}$. This makes the hierarchical collapsed configuration a promising target for CC searches, for which our analytic result provides a suitable template.

As $k_s\to k_{34}$, the nonlocal and local pieces are separately
singular, while their folded singularities cancel in the sum.
The resulting finite folded limit is visible in the left panel of
Fig.~\ref{fig:signals} and permits a smooth continuation to the
boundary $k=E$ relevant to the three-point configuration.

\emph{Three-point signal.---}
For the three-point function, we set $p=-2$ and $k=E=k_3$.
At this boundary the distinction between local and nonlocal signals
is no longer intrinsic, and the limit $k/E\to1$ is taken only after
the two contributions are combined.  Restoring the external-line
factors, we obtain the leading squeezed bispectrum
\begin{align}
 \mathfrak B_{\rm sig}
 ={}&
 \frac{\operatorname{sech}(\pi\mnu)}
 {16\pi k_{12}^6}
 \sum_{\mathsf c=\pm1}
 \left(
 \frac{k_3}{2k_{12}}
 \right)^{2\ii\mathsf c\mnu}
 \frac{2+\ii\mathsf c\mnu}
 {3+2\ii\mathsf c\mnu}
 \nonumber\\[-1mm]
 &\times
 \frac{
 \Gamma^2\!\left(\frac32+\ii\mathsf c\mnu\right)
 \Gamma(-\ii\mathsf c\mnu)}
 {\Gamma(2+\ii\mathsf c\mnu)} .
 \label{eq:bispectrum-signal}
\end{align}
The two terms are complex conjugates and produce oscillations with
frequency $2\mnu$ in $\log(k_3/k_{12})$, reflecting the two-particle
threshold of the massive loop. As illustrated in the right panel of
Fig.~\ref{fig:signals}, increasing $\mnu$ leads to more rapid
oscillations in logarithmic momentum space, while the signal amplitude
is increasingly suppressed; for $\mnu\gg1$, the suppression scales as
$e^{-2\pi\mnu}$.

%As an independent check, we have compared the analytical trispectrum and bispectrum signals with direct numerical evaluations of the corresponding loop integrals in the soft limits considered here and found excellent agreement.

As an independent check, we numerically evaluated the original uncut Schwinger–Keldysh integrals in Eqs.~\eqref{eq_Ts} and \eqref{eq_B}. The extracted four-point local signal and three-point squeezed signal show excellent agreement with the analytical results in Eqs.~\eqref{eq:trispectrum-local} and \eqref{eq:bispectrum-signal}, respectively. This provides a nontrivial numerical check of both the analytical signals and the directional cutting rule.

\emph{Discussion.---}
In this \emph{Letter}, we have obtained the complete leading CC signals from a massive scalar triangle loop in both the trispectrum and bispectrum. Our analysis employs a directional cutting rule that removes the time ordering across the cut and isolates the on-shell contributions that generate CC signals. 

Although we focus on the leading signal, subleading corrections can be computed order by order in the soft expansion, progressively extending the result away from the strict soft limit to finite momentum ratios. Importantly, our method does not rely on de Sitter boost symmetry and can therefore be applied to boost-breaking inflationary models, which are particularly relevant for generating observable CC signals. The same framework can be systematically extended to higher-spin internal states, more general interactions, and other loop topologies, opening a route toward fast and accurate analytic templates for a broader class of phenomenologically relevant loop signals.

\begin{acknowledgments}
\emph{Note added:} During the completion of this work, we learned an independent study on the numerical computation of massive scalar triangle loop process with Witten–Feynman parameterization \cite{Borinsky:2026roj}. While the two works have considered the same triangle loop bispectrum, the approaches and results are quite complementary.   

We would like to thank Dong-Gang Wang and Yi Wang for insightful discussions, Haoyuan Liu for the help with numerical checks, and Qianshu Lu for the correspondence on the numerical study of the triangle loop. 
This work is supported by NSFC under Grants No.\ 12275146 and No.\ 12247103, the National Key R\&D Program of China (2021YFC2203100), and the Dushi Program of Tsinghua University. 
\end{acknowledgments}

\bibliographystyle{utphys}
\bibliography{references}

\onecolumngrid
\begin{center}
    \textbf{End Matter}
\end{center}
\twocolumngrid

\emph{Mellin contours and residue evaluation.---} 
In this End Matter, we provide details of finishing the Mellin integrals in (\ref{eq:cut-factorization}).

By combining Eqs.~\eqref{eq:isosceles-loop} and \eqref{eq:time-part} with the PMB
Gamma factors in Eq.~\eqref{eq:cut-factorization}, we get the complete Mellin
integrand. The two terms in Eq.~\eqref{eq:isosceles-loop} are defined on the
same fundamental Mellin contours, but their expansions in the common hierarchy
$k_1\gg E\geq k$ require different contour deformations.

\emph{(i)} For $\mathbb L_1$, the relevant kinematic dependence can be written
as $(k/k_1)^{-2s_{13}}(k/E)^{-2s_{24}}$. Since $k/k_1\ll1$ while $k/E$ is
kept finite, the large-$k_1$ expansion closes the $s_{1,3}$ contours to the
left, without yet fixing the $s_{2,4}$ contours. For integer $p$, picking up the spectral poles of $\Gamma(s_{1,3}\pm\ii\mnu/2)$ is necessary for a CC signal.

\emph{(ii)} For $\mathbb L_2$, the factor $(E/k_1)^{2s_{24}}$ closes the
$s_{2,4}$ contours to the right. The poles of the time-integral factor
$\Gamma(p+4-2s_{24})$ generate only integer powers of $E$ and hence belong to
the analytic sector. At the hard-region collective poles
$s_{1234}=3/2+n$ of $\mathbb L_2$, with $n=0,1,\ldots$, however,
$(E/k_1)^{2s_{24}}=(E/k_1)^{3+2n-2s_{13}}$, which in turn closes the $s_{1,3}$
contours to the left. The same spectral-pole condition is therefore required for a CC signal.

Thus for both branches, the leading signal receives contributions only from the
same-sign spectral poles
\begin{equation}
 s_1=s_3=-\frac{\ii\mathsf c\mnu}{2},
 \qquad \mathsf c=\pm1.
\label{eq_LeadingSignalPoles}
\end{equation}
The mixed-sign choices have $s_{13}\in\mathbb Z_{\leq 0}$ and vanish for $p=0,-2$
because of the factor
$\sin[\pi(s_{13}-p/2)]$ in Eq.~\eqref{eq:time-part}.

It is therefore convenient to organize the multidimensional residues by taking
these spectral poles first. The remaining $s_{2,4}$ contours then
encounter three candidate signal-producing pole families: their individual
spectral poles
 $s_{2,4}=-n_{2,4}\pm\ii\mnu/2$; the apparent collective soft-region poles
 $t=-n$ of $\Gamma(t)$ in $\mathbb L_1$; and the apparent collective
 hard-region poles $t=n$ of $\Gamma(-t)$ in $\mathbb L_2$, where
$t\equiv s_{1234}-3/2$ and $n,n_j=0,1,\ldots$. At four points, the individual spectral residues generate the nonlocal signal, whereas the collective poles generate the local signal, as can be seen from their kinematic dependence.

Since $\mathbb L_1$ and $\mathbb L_2$ are branches of the same loop integral,
they share a single $t$ contour. Up to regular factors, their relevant collective-pole
dependence is $(k/E)^{-2t}\Gamma(t)$ and
$(E/k_1)^{2t}\Gamma(-t)$, respectively. For definiteness, we choose the common contour as $\operatorname{Re}t=\epsilon$ with $\epsilon\in(0,1)$. Then for $0<k/E<1$ and
$E/k_1\ll1$, $\mathbb L_1$ closes to the left and $\mathbb L_2$ to the right.
Therefore, this choice assigns the shared $t=0$ residue to $\mathbb L_1$, while the additional collective residues of $\mathbb L_2$ are subleading in $E/k_1$ \footnote{One may equally choose $-1<\epsilon<0$. The two prescriptions differ only in the assignment of the shared $t=0$ residue. Shifting the common contour across $t=0$ transfers this residue between the two branches, leaving their sum unchanged.}.
The remaining $s_{2,4}$ Mellin integrals are evaluated using the residue theorem and Barnes' lemma; their explicit evaluation is given in the Supplemental Material~\cite{SupplementalMaterial}. After these integrations, the
remaining $s_{5,6}$ Mellin integrals combine with the $\tau_{1,2}$
integrals to reconstruct the PMB representation of a tree-level
single-exchange seed. More explicitly, setting
$k_2=k_1$ and $s_1=s_3=-\ii\mathsf c\mnu/2$, we have
\begin{align}
 &\sum_{\mathsf a,\mathsf b=\pm}(-\mathsf a\mathsf b)
 \int_{\tau_{1,2},s_{5,6}}
 \!\!\!\!\!\!\mathcal N_{\mathsf a\mathsf b}(s_5,s_6;\tau_1,\tau_2)k_1^{-2s_{56}}e^{\mathsf a\ii k_1\tau_1+\mathsf b\ii k_1\tau_2}\nonumber\\[-1mm]
 &\times \prod_{j=5}^{6}\left[(-\tau_{j-4})^{1+\ii\mathsf c\mnu-2s_j}
 \Gamma\!\left(s_j+\frac{\ii\mnu}{2}\right)
 \Gamma\!\left(s_j-\frac{\ii\mnu}{2}\right)\right]
 \nonumber\\
  &= 
 k_1^{-4-2\ii\mathsf c\mnu}
 \mathcal I_{\mnu}^{-1/2+\ii\mathsf c\mnu,
 -1/2+\ii\mathsf c\mnu}(1,1),
 \label{eq:seed-reconstruction}
\end{align}
where $\mathcal I^{p_1,p_2}_{\mnu}$ denotes the tree-level single-exchange seed, whose general-index expression was obtained in Ref.~\cite{Qin:2023ejc}. For the particular indices appearing in Eq.~\eqref{eq:seed-reconstruction}, Ref.~\cite{Qin:2023bjk} gives
\begin{align}
  &\mathcal I_{\mnu}^{-1/2+\ii\mathsf c\mnu,
 -1/2+\ii\mathsf c\mnu}(1,1)\nonumber\\
 &\quad=-\frac{(2+\ii\mathsf c\mnu)
\sin(\ii\mathsf c\pi\mnu)}
{\sqrt{\pi}\,(3+2\ii\mathsf c\mnu)}
\Gamma(1+\ii\mathsf c\mnu)
\Gamma\!\left(\frac32+\ii\mathsf c\mnu\right).
 \label{eq:seed-special}
\end{align}

%Denoting the result of the remaining $s_{2,4}$ integrations by
%$\mathcal M_{\mathsf c}^p(k/E)$, whose explicit hypergeometric form is
%given in the Supplemental Material~\cite{SupplementalMaterial}, the
%leading master signal factorizes as
%\begin{align}
% &\mathcal T_{\rm sig}^p(\bm k,E)
% =
% \frac{k^3/(k_1^4E^{p+4})}{64\pi^{7/2}}
% \sum_{\mathsf c=\pm1}
% \left(\frac{k}{2k_1}\right)^{2\ii\mathsf c\mnu}
% \mathcal M_{\mathsf c}^p\!\left(\frac{k}{E}\right)\nonumber\\
% &\times \sin\!\left[
% \pi\!\left(\ii\mathsf c\mnu+\fr p2\right)\right]
% \Gamma^2(-\ii\mathsf c\mnu)
% \,\mathcal I_{\mnu}^{-\frac12+\ii\mathsf c\mnu,
% -\frac12+\ii\mathsf c\mnu}(1,1).
% \label{eq:master-signal}
%\end{align}

\clearpage
\onecolumngrid

\begin{center}
 {\large\bfseries Supplemental Material for}\\[0.4em]
 {\large\bfseries\textit{Cosmological Collider Signals From a Triangle Loop}}\\[1em]
 Zhehan Qin$^{1,2}$ and Zhong-Zhi Xianyu$^{1,3}$\\
 \textit{${}^1$~Department of Physics, Tsinghua University, Beijing 100084, China}\\
 \textit{${}^2$~Department of Applied Mathematics and Theoretical Physics,\\University of Cambridge, Wilberforce Road, Cambridge, CB3 0WA, UK}\\
 \textit{${}^3$~Peng Huanwu Center for Fundamental Theory, Hefei, Anhui 230026, China}\\
 \texttt{qzh17@tsinghua.org.cn},
 \texttt{zxianyu@tsinghua.edu.cn}
\end{center}

\setcounter{equation}{0}
\setcounter{figure}{0}
\setcounter{table}{0}
\renewcommand{\theequation}{S\arabic{equation}}
\renewcommand{\thefigure}{S\arabic{figure}}
\renewcommand{\thetable}{S\arabic{table}}

In this Supplemental Material, we follow the notation and conventions
of the main text.  We derive the directional cutting rule, evaluate
the isosceles massless triangle integral exactly, and give details of
the remaining Mellin integrations.

\section{Propagator conventions}

For completeness, we collect here the propagator conventions used in
the main text.  The bulk-to-boundary propagator of the inflaton
fluctuation and its conformal-time derivative are
\begin{align}
K_\pm(k,\tau)
&=
\frac{(1\mp i k\tau)e^{\pm i k\tau}}{2k^3},
&
K_\pm'(k,\tau)
&=
\frac{\tau}{2k}e^{\pm i k\tau}.
\end{align}
For a general massive field lying in the principal series, the bulk-to-bulk propagators are
\begin{align}
&D_{-+}(k;\tau_1,\tau_2)
=
\sigma_k(\tau_1)\sigma_k^*(\tau_2),\qquad
D_{+-}(k;\tau_1,\tau_2)
=
\sigma_k^*(\tau_1)\sigma_k(\tau_2),\nonumber\\
&D_{\pm\pm}(k;\tau_1,\tau_2)
=
\theta(\tau_1-\tau_2)D_{\mp\pm}(k;\tau_1,\tau_2)
+
\theta(\tau_2-\tau_1)D_{\pm\mp}(k;\tau_1,\tau_2),
\end{align}
with $\sigma_k(\tau)$ the corresponding mode function,
\begin{equation}
  \sigma_k(\tau) =\frac{\sqrt{\pi}}{2}
 e^{-\pi\mnu/2}(-\tau)^{3/2}
 \mathrm H_{\ii\mnu}^{(1)}(-k\tau).
\end{equation}

\section{Derivation of the directional cutting rule}
\label{sec:directional-cut}

In this section we derive the directional cutting rule used in the
main text directly from the full master integral.  We work first in
the hierarchy
\begin{equation}
 k_1=k_2=\frac{k_{12}}{2}\gg E>k>0,
 \label{eq:sup-hierarchy}
\end{equation}
and show that all cosmological collider signals of the full master
integral are contained in the directional-cut contribution.  We use
$\{\tau\}\equiv(\tau_1,\tau_2,\tau_3)$ and
$\{s\}\equiv(s_1,\ldots,s_6)$, as in the main text.  The four-point
configuration has $(\bm k,E)=(\bm k_s,k_{34})$, whereas the three-point
configuration has $(\bm k,E)=(-\bm k_3,k_3)$.  The boundary $E=k$, which
includes both the folded endpoint of the four-point configuration and the
three-point configuration, is obtained by analytic continuation from
$0<k/E<1$ after the relevant signal branches have been combined.  The
argument below is needed for $p=0,-2$ and is understood with the standard
Bunch--Davies $\ii\epsilon$ prescription, initially in a common convergence
domain and subsequently by analytic continuation.

\paragraph{Full PMB decomposition.---}
Applying the PMB representation to all three massive propagators in
the full master integral gives
\begin{align}
 \mathcal T^p(\bm k,E)
 ={}&\int_{\{s\}}
 \mathbb T^p(\{s\};k_1,k_2,E)
 \mathbb L(\{s\};\bm k,\bm k_1)
 \prod_{j=1}^{6}
 \Gamma\!\left(s_j+\frac{\ii\mnu}{2}\right)
 \Gamma\!\left(s_j-\frac{\ii\mnu}{2}\right),
 \label{eq:sup-full-factorization}
\end{align}
where
\begin{equation}
 \mathbb L
 =
 \int\frac{\dd^3\bm q}{(2\pi)^3}
 q^{-2s_{12}}
 |\bm q+\bm k|^{-2s_{34}}
 |\bm q+\bm k_1|^{-2s_{56}},
 \label{eq:sup-loop}
\end{equation}
contains all loop-momentum dependence, while
\begin{align}
 \mathbb T^p
 ={}&
 \sum_{\mathsf a,\mathsf b,\mathsf c=\pm}
 (-\ii\mathsf a\mathsf b\mathsf c)
 \int_{\{\tau\}}
 \mathcal N_{\mathsf a\mathsf c}
 (s_1,s_2;\tau_1,\tau_3)
 \mathcal N_{\mathsf b\mathsf c}
 (s_3,s_4;\tau_2,\tau_3)
 \mathcal N_{\mathsf a\mathsf b}
 (s_5,s_6;\tau_1,\tau_2)\nonumber\\
 &\times
 (-\tau_1)^{1-2s_{15}}
 (-\tau_2)^{1-2s_{36}}
 (-\tau_3)^{p+3-2s_{24}}
 e^{\ii\mathsf a k_1\tau_1
 +\ii\mathsf b k_2\tau_2
 +\ii\mathsf c E\tau_3},
 \label{eq:sup-full-time}
\end{align}
contains all Schwinger--Keldysh time ordering.

At fixed hard momenta, all explicit dependence on the soft energy $E$
is carried by the time part $\mathbb T^p$, whereas all explicit dependence on
the soft momentum $k$ resides in the loop part $\mathbb L$.  In the
configurations considered here, the only nonanalyticities
relevant to the cosmological collider signal therefore fall into two classes:
the soft-energy nonanalyticity in $E/k_1$ and the soft-momentum nonanalyticity
in $k/k_1$.  For the four-point configuration
$(E,k)=(k_{34},k_s)$, the soft-energy nonanalyticity corresponds to the local
signal, while the soft-momentum
nonanalyticity corresponds to the nonlocal signal.  For the three-point
configuration $E=k=k_3$, the two soft variables coincide, so this distinction
is no longer intrinsic and the two contributions combine into the complete
squeezed signal.

The two nesting factors associated with the lines ending at
$\tau_3$ can be decomposed exactly as
\begin{align}
 \mathcal N_{\mathsf a\mathsf c}
 (s_1,s_2;\tau_1,\tau_3)
 ={}&
 \mathcal N_{-\mathsf c,\mathsf c}(s_1,s_2)
 +\delta_{\mathsf a,\mathsf c}\,
 \theta_{31}\,
 \Delta\mathcal N_{\mathsf c}(s_1,s_2),
 \label{eq:sup-N-decomp1}\\
 \mathcal N_{\mathsf b\mathsf c}
 (s_3,s_4;\tau_2,\tau_3)
 ={}&
 \mathcal N_{-\mathsf c,\mathsf c}(s_3,s_4)
 +\delta_{\mathsf b,\mathsf c}\,
 \theta_{32}\,
 \Delta\mathcal N_{\mathsf c}(s_3,s_4),
 \label{eq:sup-N-decomp2}
\end{align}
where
\begin{align}
 \Delta\mathcal N_{\mathsf c}(s_i,s_j)
 &\equiv
 \mathcal N_{\mathsf c,-\mathsf c}(s_i,s_j)
 -
 \mathcal N_{-\mathsf c,\mathsf c}(s_i,s_j)=
 -\frac{\ii\mathsf c\,2^{2s_{ij}}}{2\pi}
 \sin\!\left[\pi(s_i-s_j)\right].
 \label{eq:sup-deltaN}
\end{align}
Substituting Eqs.~\eqref{eq:sup-N-decomp1} and
\eqref{eq:sup-N-decomp2} into Eq.~\eqref{eq:sup-full-time}
decomposes the full time integral into
\begin{equation}
 \mathbb T^p
 =
 \mathbb T_{\mathrm{cut}}^p
 +\mathbb R_1^p
 +\mathbb R_2^p
 +\mathbb R_{12}^p.
 \label{eq:sup-time-decomposition}
\end{equation}
Here the cut piece $\mathbb T_{\mathrm{cut}}^p$ contains the product $\mathcal N_{-\mathsf c,\mathsf c}(s_1,s_2)
 \mathcal N_{-\mathsf c,\mathsf c}(s_3,s_4)$,
while the remainder terms $\mathbb R_1^p$, $\mathbb R_2^p$, and
$\mathbb R_{12}^p$ contain respectively $\theta_{31}$,
$\theta_{32}$, and $\theta_{31}\theta_{32}$.  We now show that these three
remainder terms generate neither of the two signal nonanalyticities identified
above.

\paragraph{Soft-energy nonanalyticity.---}
We first consider possible nonanalytic dependence on the soft energy
$E$. Every remainder term in Eq.~\eqref{eq:sup-time-decomposition}
contains at least one step function $\theta_{31}$ or $\theta_{32}$.
For example, $\theta_{31}$ restricts $\tau_3$ to
$\tau_1<\tau_3<0$. Writing $\beta=p+3-2s_{24}$, the corresponding
ordered integral is
\begin{align}
 \int_{\tau_1}^{0}\dd\tau_3\,
 (-\tau_3)^\beta e^{\ii\mathsf cE\tau_3}
 =
 \sum_{n=0}^{\infty}
 \frac{(-\ii\mathsf cE)^n}
 {n!(\beta+n+1)}
 (-\tau_1)^{\beta+n+1}.
\end{align}
Termwise integration, performed first in the common convergence
domain, therefore produces only integer powers of $E/k_1$.
The same argument applies to $\mathbb R_2^p$ and
$\mathbb R_{12}^p$. Hence none of the remainder terms can generate
the mass-dependent complex-power nonanalyticity in the soft energy.

\paragraph{Soft-momentum nonanalyticity.---}
It remains to check for possible soft-momentum nonanalyticity from the
loop factor, which can arise only from the first term of the exact
triangle integral in Eq.~\eqref{eq:supp-triangle-exact}. We denote this
term by $\mathbb L_1$. Up to factors regular at the spectral poles, its
relevant dependence is
\begin{equation}
 \mathbb L_1 \propto k^{3-2s_{1234}}
 \frac{
 \Gamma\!\left(s_{1234}-\frac32\right)
 \Gamma\!\left(\frac32-s_{12}\right)
 \Gamma\!\left(\frac32-s_{34}\right)
 }{
 \Gamma(s_{12})\Gamma(s_{34})
 },
 \label{eq:sup-L1-relevant}
\end{equation}
and the oscillatory $k$ dependence requires spectral poles of
the Mellin variables associated with the two soft internal lines,
\begin{equation}
 s_j=-n_j-\frac{\ii\mathsf c_j\mnu}{2},
 \qquad
 \mathsf c_j=\pm1,
 \qquad
 n_j=0,1,\ldots,
 \qquad j=1,\ldots,4.
 \label{eq:sup-spectral-poles}
\end{equation}

Consider first a remainder containing
$\Delta\mathcal N_{\mathsf c}(s_1,s_2)$.  If the two spectral poles
have the same sign, $\mathsf c_1=\mathsf c_2$, then $s_1-s_2=n_2-n_1\in\mathbb Z$, and Eq.~\eqref{eq:sup-deltaN} gives $\Delta\mathcal N_{\mathsf c}(s_1,s_2)=0$.
Thus the corresponding spectral residue vanishes.
If instead the two spectral poles have opposite signs,
$\mathsf c_1=-\mathsf c_2$, then
$s_{12}=-(n_1+n_2)\in\mathbb Z_{\leq0}$,
and the loop factor in Eq.~\eqref{eq:sup-L1-relevant} vanishes because $1/\Gamma(s_{12})=0$.
Hence the mixed-sign spectral residue vanishes as well.

Exactly the same argument applies to a remainder containing
$\Delta\mathcal N_{\mathsf c}(s_3,s_4)$, using the factor
$1/\Gamma(s_{34})$ in Eq.~\eqref{eq:sup-L1-relevant}.  Since every
non-cut contribution in Eq.~\eqref{eq:sup-time-decomposition}
contains at least one of these two $\Delta\mathcal N$ factors, all
spectral residues that could generate the nonlocal loop signal
vanish in $\mathbb R_1^p$, $\mathbb R_2^p$, and
$\mathbb R_{12}^p$.

Combining the two observations, the remainder terms generate
neither type of signal nonanalyticity.
Therefore, in the hierarchy of Eq.~\eqref{eq:sup-hierarchy},
the complete cosmological collider signal of the full master
integral is contained in the directional-cut piece:
\begin{equation}
 \mathcal T_{\mathrm{sig}}^p(\bm k,E)
 =
 \mathcal T_{\mathrm{cut,sig}}^p(\bm k,E).
 \label{eq:sup-cut-equality}
\end{equation}
This establishes the directional cutting rule used in the main text.  The
$E=k$ configurations follow by analytic continuation of the combined signal
from $0<k/E<1$ to the boundary $k/E=1$.

\section{Isosceles massless triangle integral}
\label{sec:isosceles-triangle}

We derive here the exact isosceles massless triangle integral whose leading small-$k/k_{12}$ form is used in the main text.  Writing
$a=s_{12}$, $b=s_{34}$, and $c=s_{56}$, we temporarily regard
$a,b,c$ as independent complex parameters and define
\begin{equation}
 \mathbb L(a,b,c;\bm k,\bm k_1)
 \equiv
 \int\frac{\dd^3\bm q}{(2\pi)^3}
 q^{-2a}|\bm q+\bm k|^{-2b}|\bm q+\bm k_1|^{-2c}.
 \label{eq:supp-triangle-definition}
\end{equation}
We suppress the momentum arguments below.  All formulas are first
understood in a common domain of absolute convergence and then
continued meromorphically in $a,b,c$.

Introducing Feynman parameters and shifting the loop momentum gives
\begin{align}
 \mathbb L(a,b,c)
 ={}&\frac{1}{(4\pi)^{\frac32}}
 \frac{\Gamma\!\left(A-\frac32\right)}
 {\Gamma(a)\Gamma(b)\Gamma(c)}
 \int_0^1\prod_{i=1}^3\dd\xi_i\,
 \delta(1-\xi_1-\xi_2-\xi_3)
 \xi_1^{a-1}\xi_2^{b-1}\xi_3^{c-1}
 \Delta^{\frac32-A},
 \label{eq:supp-triangle-parameters}
\end{align}
where $A\equiv a+b+c$ and $\Delta = \xi_1\xi_2 k^2
 +\xi_1\xi_3 k_1^2
 +\xi_2\xi_3 k_2^2$.
For the isosceles configuration relevant to the leading limit in
the main text, $k_1=k_2=k_{12}/2$, the quadratic form reduces to
\begin{equation}
 \Delta=\xi_1\xi_2k^2+\xi_3(1-\xi_3)k_1^2.
\end{equation}
We separate the two terms using the Mellin--Barnes identity
\begin{equation}
 \frac{1}{(X+Y)^\lambda}
 =
 \int_{-\ii\infty}^{+\ii\infty}\frac{\dd z}{2\pi\ii}\,
 X^{-\lambda-z}Y^z
 \frac{\Gamma(\lambda+z)\Gamma(-z)}
 {\Gamma(\lambda)}.
 \label{eq:supp-mb-identity}
\end{equation}
Taking
\begin{equation}
 X=\xi_1\xi_2k^2,\qquad
 Y=\xi_3(1-\xi_3)k_1^2,\qquad
 \lambda=A-\frac32,
\end{equation}
and performing the remaining Feynman-parameter integrals by Beta
integrals gives the one-fold Mellin--Barnes representation
\begin{align}
 \left.\mathbb L(a,b,c)\right|_{k_1=k_2}
 ={}&\frac{k^{3-2a-2b-2c}}{(4\pi)^{\frac32}}
 \int_{-\ii\infty}^{+\ii\infty}\frac{\dd z}{2\pi\ii}
 \left(\frac{k}{k_1}\right)^{-2z}
 \frac{\Gamma\!\left(a+b+c-\frac32+z\right)\Gamma(-z)}
 {\Gamma(a)\Gamma(b)\Gamma(c)}
 \nonumber\\[-1mm]
 &\times
 \frac{
 \Gamma\!\left(\frac32-a-c-z\right)
 \Gamma\!\left(\frac32-b-c-z\right)
 \Gamma(c+z)
 \Gamma(3-a-b-2c-z)}
 {\Gamma(3-a-b-2c-2z)
 \Gamma(3-a-b-c)},
 \label{eq:supp-triangle-mb}
\end{align}
where the contour is chosen to separate the left-pole families
\begin{equation}
 z=\frac32-a-b-c-n,
 \qquad
 z=-c-n,
 \qquad
 n=0,1,\ldots,
 \label{eq:supp-z-left-poles}
\end{equation}
from the right-pole families
\begin{align}
 z=n,\qquad
 z=\frac32-a-c+n,\qquad
 z=\frac32-b-c+n,\qquad
 z=3-a-b-2c+n,\qquad n=0,1,\ldots.
 \label{eq:supp-z-right-poles}
\end{align}

For $0<k<2k_1$, the contour is closed to the left.  Summing the two
residue families in Eq.~\eqref{eq:supp-z-left-poles} and defining
\begin{equation}
 x\equiv\frac{k^2}{4k_1^2}=\frac{k^2}{k_{12}^2},
\end{equation}
we obtain the exact isosceles result
\begin{align}
 \left.\mathbb L(a,b,c)\right|_{k_1=k_2}
 ={}&\frac{k_1^{-2c}k^{3-2a-2b}}{(4\pi)^{\frac32}}
 \frac{
 \Gamma\!\left(a+b-\frac32\right)
 \Gamma\!\left(\frac32-a\right)
 \Gamma\!\left(\frac32-b\right)}
 {\Gamma(3-a-b)\Gamma(a)\Gamma(b)}
 \nonumber\\[-1mm]
 &\times
 {}_4\mathrm F_3\!\left[
 \begin{matrix}
 3-a-b-c,\;\frac32-a,\;\frac32-b,\;c\\
 \frac52-a-b,\;\frac32-\frac{a+b}{2},\;2-\frac{a+b}{2}
 \end{matrix}
 \middle|x
 \right]
 \nonumber\\[1mm]
 &+\frac{k_1^{3-2a-2b-2c}}{(4\pi)^{\frac32}}
 \frac{
 \Gamma\!\left(a+b+c-\frac32\right)
 \Gamma\!\left(\frac32-a-b\right)
 \Gamma\!\left(\frac32-c\right)}
 {\Gamma(3-a-b-c)\Gamma(a+b)\Gamma(c)}
 \nonumber\\[-1mm]
 &\times
 {}_4\mathrm F_3\!\left[
 \begin{matrix}
 a+b+c-\frac32,\;a,\;b,\;\frac32-c\\
 a+b-\frac12,\;\frac{a+b}{2},\;\frac{a+b+1}{2}
 \end{matrix}
 \middle|x
 \right].
 \label{eq:supp-triangle-exact}
\end{align}

The two hypergeometric branches resum the two pole families in
Eq.~\eqref{eq:supp-z-left-poles}.  Their series converge throughout the nonfolded isosceles range $0<x<1$, while the folded point $x=1$
is obtained as a boundary limit.  Importantly, the two branches are
not separate loop integrals but inherit the same original Mellin
contours.
Restoring $(a,b,c)=(s_{12},s_{34},s_{56})$ and expanding for
$x\ll1$, as appropriate to the hierarchy in the main text, we have
${}_4\mathrm F_3(\cdots|x)=1+\order{x}$, and
Eq.~\eqref{eq:supp-triangle-exact} gives
\begin{align}
 \mathbb L
 ={}&
 \frac{k_1^{-2s_{56}}k^{3-2s_{1234}}}{(4\pi)^{\frac32}}
 \frac{
 \Gamma\!\left(s_{1234}-\frac32\right)
 \Gamma\!\left(\frac32-s_{12}\right)
 \Gamma\!\left(\frac32-s_{34}\right)}
 {\Gamma(3-s_{1234})
 \Gamma(s_{12})
 \Gamma(s_{34})}
 \left[1+\order{x}\right]
 \nonumber\\[1mm]
 &+
 \frac{k_1^{3-2s_{123456}}}{(4\pi)^{\frac32}}
 \frac{
 \Gamma\!\left(s_{123456}-\frac32\right)
 \Gamma\!\left(\frac32-s_{1234}\right)
 \Gamma\!\left(\frac32-s_{56}\right)}
 {\Gamma(3-s_{123456})
 \Gamma(s_{1234})
 \Gamma(s_{56})}
 \left[1+\order{x}\right].
 \label{eq:supp-triangle-small-k}
\end{align}
The leading terms reproduce the result quoted in the main text.

\section{Evaluation of the remaining Mellin integrals}
\label{sec:remaining-mellin}

We now evaluate explicitly the remaining $s_{2,4}$ Mellin integrals
left implicit in the main text.  After taking the leading same-sign
spectral poles
\begin{equation}
 s_1=s_3=-\frac{\ii\mathsf c\mnu}{2},
 \qquad
 \mathsf c=\pm1,
 \label{eq:supp-s13-poles}
\end{equation}
For the common contour choice
$\operatorname{Re}t=\epsilon$ with $0<\epsilon<1$, the shared
$t=0$ contribution is assigned to the first term of the loop
integral.  The additional collective poles of the second term are
suppressed by powers of $E/k_1$, while its time-integral poles belong
to the analytic sector.  Hence the leading signal can be obtained
from the first term of the small-$k$ expansion.

After taking Eq.~\eqref{eq:supp-s13-poles}, cancelling the Gamma
functions common to the PMB factors and the denominator of the loop
integral, and factoring out all terms independent of $s_{2,4}$, the
remaining two-fold integral can be written as
\begin{align}
 \mathcal M_{\mathsf c}^p\!\left(\frac{k}{E}\right)
 \equiv{}&
 \int_{s_{2,4}}
 \left(\frac{k}{2E}\right)^{-2s_{24}}
 \frac{
 \Gamma(p+4-2s_{24})
 \Gamma\!\left(s_{24}-\frac32-\ii\mathsf c\mnu\right)}
 {\Gamma(3-s_{24}+\ii\mathsf c\mnu)}
 \prod_{j=2,4}
 \Gamma\!\left(s_j+\frac{\ii\mathsf c\mnu}{2}\right)
 \Gamma\!\left(\frac32-s_j+\frac{\ii\mathsf c\mnu}{2}\right).
 \label{eq:supp-reduced-mellin}
\end{align}
For $0<k<E$, the $s_2$ and $s_4$ contours are closed to the left.
The relevant enclosed poles are either the individual spectral poles
or the collective soft-region loop-UV poles
$s_{24}-3/2-\ii\mathsf c\mnu=-n$; these generate the two signal
contributions below, respectively.

\paragraph{Nonlocal signal.---}
The individual spectral poles are
\begin{equation}
 s_2=-n_2-\frac{\ii\mathsf c\mnu}{2},
 \qquad
 s_4=-n_4-\frac{\ii\mathsf c\mnu}{2},
 \qquad
 n_2,n_4=0,1,\ldots .
 \label{eq:supp-individual-poles}
\end{equation}
Writing $N=n_2+n_4$, the sum over partitions of $N$ is simplified by
\begin{equation}
 \sum_{n_2+n_4=N}
 \frac{
 (\frac32+\ii\mathsf c\mnu)_{n_2}
 (\frac32+\ii\mathsf c\mnu)_{n_4}}
 {n_2!\,n_4!}
 =
 \frac{(3+2\ii\mathsf c\mnu)_N}{N!}.
 \label{eq:supp-vandermonde}
\end{equation}
The resulting residue sum gives
\begin{align}
 \mathcal M_{\mathsf c,\mathrm{NS}}^p\!\left(\frac{k}{E}\right)
 ={}&
 \left(\frac{k}{2E}\right)^{2\ii\mathsf c\mnu}
 \frac{
 \Gamma^2\!\left(\frac32+\ii\mathsf c\mnu\right)
 \Gamma\!\left(-\frac32-2\ii\mathsf c\mnu\right)
 \Gamma(p+4+2\ii\mathsf c\mnu)}
 {\Gamma(3+2\ii\mathsf c\mnu)}
 {}_2\mathrm F_1\!\left[
 \begin{matrix}
 2+\frac p2+\ii\mathsf c\mnu,\;
 \frac52+\frac p2+\ii\mathsf c\mnu\\
 \frac52+2\ii\mathsf c\mnu
 \end{matrix}
 \middle|\frac{k^2}{E^2}
 \right].
 \label{eq:supp-spectral-result}
\end{align}
For the four-point function this contribution gives the nonlocal
signal.

\paragraph{Local signal.---}
The collective poles of the first term are
\begin{equation}
 s_{24}=\frac32+\ii\mathsf c\mnu-n,
 \qquad
 n=0,1,\ldots .
 \label{eq:supp-collective-poles}
\end{equation}
At each such pole one Mellin integration remains.  Setting
$s_4=3/2+\ii\mathsf c\mnu-n-s_2$, the relevant Barnes integral is
\begin{align}
 &\int_{s_2}
 \Gamma\!\left(s_2+\frac{\ii\mathsf c\mnu}{2}\right)
 \Gamma\!\left(s_2+n-\frac{\ii\mathsf c\mnu}{2}\right)
 \Gamma\!\left(\frac32-s_2+\frac{\ii\mathsf c\mnu}{2}\right)
 \Gamma\!\left(\frac32+\frac{3\ii\mathsf c\mnu}{2}-n-s_2\right)
 \nonumber\\
 ={}&
 \frac{
 \Gamma^2\!\left(\frac32+\ii\mathsf c\mnu\right)
 \Gamma\!\left(\frac32+2\ii\mathsf c\mnu-n\right)
 \Gamma\!\left(\frac32+n\right)}
 {\Gamma(3+2\ii\mathsf c\mnu)}.
 \label{eq:supp-barnes-integral}
\end{align}
Combining this with the residue of
$\Gamma(s_{24}-3/2-\ii\mathsf c\mnu)$ and summing over $n$ gives
\begin{align}
 \mathcal M_{\mathsf c,\mathrm{LS}}^p\!\left(\frac{k}{E}\right)
 ={}&
 \left(\frac{k}{2E}\right)^{-3-2\ii\mathsf c\mnu}
 \frac{
 \Gamma(p+1-2\ii\mathsf c\mnu)
 \Gamma^2\!\left(\frac32+\ii\mathsf c\mnu\right)
 \Gamma\!\left(\frac32+2\ii\mathsf c\mnu\right)}
 {\Gamma(3+2\ii\mathsf c\mnu)}
 {}_2\mathrm F_1\!\left[
 \begin{matrix}
 \frac{p+1}{2}-\ii\mathsf c\mnu,\;
 \frac{p+2}{2}-\ii\mathsf c\mnu\\
 -\frac12-2\ii\mathsf c\mnu
 \end{matrix}
 \middle|\frac{k^2}{E^2}
 \right].
 \label{eq:supp-collective-result}
\end{align}
At four points this is the local signal generated by the collective
pole family.

The complete leading $s_{2,4}$ contribution is therefore
\begin{equation}
 \mathcal M_{\mathsf c}^p\!\left(\frac{k}{E}\right)
 =
 \mathcal M_{\mathsf c,\mathrm{NS}}^p\!\left(\frac{k}{E}\right)
 +
 \mathcal M_{\mathsf c,\mathrm{LS}}^p\!\left(\frac{k}{E}\right).
 \label{eq:supp-mellin-result}
\end{equation}

For generic $p$ with $\operatorname{Re}p<-2$, both hypergeometric
functions converge at $k/E=1$, since their parametric excess is
$c-a-b=-2-p$.  Gauss's theorem then gives
\begin{align}
 \mathcal M_{\mathsf c,\mathrm{NS}}^p(1)
 ={}&
 \frac{
 \pi\,2^{-3-p-2\ii\mathsf c\mnu}
 \Gamma(-2-p)
 \Gamma\!\left(\frac32+\ii\mathsf c\mnu\right)
 \Gamma(4+p+2\ii\mathsf c\mnu)}
 {\cos(2\pi\ii\mathsf c\mnu)
 \Gamma(2+\ii\mathsf c\mnu)
 \Gamma(-p+2\ii\mathsf c\mnu)},\\
 \mathcal M_{\mathsf c,\mathrm{LS}}^p(1)
 ={}&-
 \frac{
 \pi\,2^{-3-p-2\ii\mathsf c\mnu}
 \Gamma(-2-p)
 \Gamma(1+p-2\ii\mathsf c\mnu)
 \Gamma\!\left(\frac32+\ii\mathsf c\mnu\right)}
 {\cos(2\pi\ii\mathsf c\mnu)
 \Gamma(-3-p-2\ii\mathsf c\mnu)
 \Gamma(2+\ii\mathsf c\mnu)}.
 \label{eq:supp-unit-argument-branches}
\end{align}
Although the two terms separately develop singularities upon
continuation to $p=0,-2$, these singularities cancel in their sum.
Using the reflection formula, we obtain
\begin{equation}
 \mathcal M_{\mathsf c}^{p}(1)
 =
 \frac{
 \pi\,2^{-2-p-2\ii\mathsf c\mnu}
 \Gamma\!\left(\frac32+\ii\mathsf c\mnu\right)
 \Gamma(4+p+2\ii\mathsf c\mnu)
 \Gamma(1+p-2\ii\mathsf c\mnu)}
 {\Gamma(3+p)\Gamma(2+\ii\mathsf c\mnu)}.
 \label{eq:supp-unit-argument-mellin}
\end{equation}
The last expression provides the analytic continuation to the physical
values $p=0,-2$.
These expressions give the complete $s_{2,4}$-dependent part of the
leading master signal.  The remaining $s_{5,6}$ Mellin integrals and
the $\tau_{1,2}$ integrations reconstruct the tree-level
single-exchange seed used in the main text.

\end{document}